\documentstyle[epsf]{ptptex}

\def\Journal#1#2#3#4{{#1}{\bf #2} (#4), #3}
\def\PTP{Prog.~Theor.~Phys.~}
\def\PTPS{Prog.~Theor.~Phys.~Suppl.~}

\def\NPA{Nucl.~Phys. \bf{A}}
\def\NPB{Nucl.~Phys. \bf{B}}

\def\PL{Phys.~Lett.~}
\def\PRL{Phys.~Rev.~Lett.~}
\def\PRD{Phys.~Rev. \bf{D}}
\def\ZPC{Z.~Phys. \bf{C}}
\notypesetlogo  
\title{%
Analysis of $K\pi$-scattering phase shift \\
and
existence of $\kappa$(900)-particle
}

\author{
Shin {\sc Ishida}, 
Muneyuki {\sc Ishida}$^{*}$,
Taku {\sc Ishida}$^{**}$\\
Kunio {\sc Takamatsu}$^{***}$,
and Tsuneaki {\sc Tsuru}$^{**}$
}

\inst{%
Atomic Energy Research Institute, 
College of Science and Technology,\\
Nihon University, Tokyo 101
\\
$^*$Department of Physics, 
University of Tokyo, Tokyo 113
\\
$^{**}$National Laboratory for High Energy Physics (KEK), 
Tsukuba 305
\\
$^{***}$Department of Engineering, Miyazaki University, 
Miyazaki 988-12
}

\recdate{%
February 1, 1997
}

\abst{%
Recently we have shown an evidence for existence
of $\sigma$-particle in the previous works; where
the $\pi\pi$ S-wave phase shift is reanalyzed, by
introducing a repulsive background suggested by
the chiral symmetry, and by applying a new method
of Interfering Breit-Wigner Amplitudes.
In this work we also show, reanalyzing the $K\pi$
S-wave phase shift from a similar standpoint,
an evidence for existence of $\kappa$(900),
possibly to be a member of $\sigma$-nonet.
}

\begin{document}

\maketitle

\section{Introduction}\label{sec:int}
In our previous works\cite{rf:pipip,rf:pl} we have analyzed
the iso-singlet S-wave $\pi\pi$ scattering phase shift,
and have shown the existence of a resonance with mass of 535-650 MeV and
width of about 350 MeV. These values are 
consistent with those of $\sigma$ particle,
the long-sought chiral partner of Nambu-Goldstone $\pi$ meson,
predicted by the linear $\sigma$ model.\cite{rf:MY} \  
Independent analyses of the phase shift 
by several authors\cite{rf:Ka,rf:Tor,rf:Ha}
have also suggested its existence.
On the other hand anticipation for $\sigma$ existence has been
given recently with new interests both
\cite{rf:sca2}\tocite{rf:Taku} 
theoretically 
and phenomenologically.
As a matter of fact a low-mass isoscalar resonance, 
$f_0$(400$\sim$1200) or $\sigma$, has revived in the latest issue of 
Particle Data Group\cite{rf:PDG96} after its missing over twenty years.

In the phase shift analysis\cite{rf:pipip,rf:pl} on one hand,
we have developed a new method of S-matrix parametrization 
in conformity with unitarity,
Interfering Breit-Wigner Amplitude method, in which we use only a few 
parameters with direct
physical meaning ({\em i.e.} masses and widths of resonances), 
in contrast with the conventional ${\cal K}$-matrix method. 
On the other hand we have introduced an negative background phase
$\delta^{BG}$ of hard core type phenomenologically.
This type of background phase shifts\cite{rf:core} was observed historically
in the $\alpha$-$\alpha$ scattering, 
and also in the nucleon-nucleon scattering.
In the relevant case of $\pi\pi$ system its origin 
seems to have some correspondence\cite{rf:MY} to 
the ``compensating" contact $\lambda\phi^4$ term required by the
chiral symmetry in the linear $\sigma$ model.

The $\sigma$-particle is a chiral partner of $\pi$-meson 
in the linear representation
of chiral $SU(2)_L\times SU(2)_R$ group. 
Taking $SU(3)$ flavor symmetry into account, 
it is natural to expect existence of scalar $\sigma$-meson nonet
as a chiral partner of pseudoscalar $\pi$-meson nonet.
In the following we analyze the I=1/2 $K\pi$ scattering 
phase shift from a similar 
standpoint to the $\pi\pi$ system, 
and actually show an evidence for existence of
I=1/2 member of the $\sigma$-nonet, $\kappa$-meson.

\section{Applied formulas}
\label{sec:app}
We analyze the I=1/2 S-wave phase shift of $K\pi$-scattering
by Interfering Amplitude method in the case
of three-channels ($K\pi ,K\eta$ and $K\eta^\prime$,
denoted to $1,2$ and $3$, respectively) 
with two-resonances ($\kappa$ and $K_0^*(1430)$) 
from the $K\pi$-threshold to $\sqrt{s}\sim 1.6$ GeV.\footnote{
Contributions from the 
other channels such as $K\pi\sigma$ and $K\pi\pi\pi$ are
expected to be supressed by a phase space factor.
}

The relevant S-matrix element\footnote{ 
The other elements of $S_{ij}(i\neq 1$ and/or $j\neq 1)$ are now irrelevant
since of present experimental situations in the corresponding processes.}
of $K\pi$-scattering, $S_{11}$, 
is related to its phase shift
(amplitude), $\delta_0^\frac{1}{2}$($a_0^\frac{1}{2}$), by
\begin{eqnarray} 
S_{11} = \eta_{11}e^{2i\delta_0^\frac{1}{2}}=1+2ia_0^\frac{1}{2},
\label{eq:Sda}
\end{eqnarray}
where $\eta_{11}$ is the elasticity.
$S_{11}$ is given by product of
``individual'' resonance-S-matrices $S^{(R)}_{11}(R=\kappa ,K_0^*)$ 
\begin{eqnarray}
S_{11} = e^{2i\delta^{BG}}\prod_{R=\kappa ,K_0^*}S^{(R)}_{11};
\label{eq:IAM}
\end{eqnarray}
The unitarity of ``total'' S-matrix (\ref{eq:IAM}) is now easily seen to be 
satisfied by the unitarity of individual S matrices.
In Eq.(\ref{eq:IAM}) we have also introduced 
a negative background phase $\delta^{BG}$,
taken as a hard core type phenomenologically,
\begin{eqnarray}
\delta^{BG} = -|{\bf p}_1|r_{c0}^\frac{1}{2};\ \ 
|{\bf p}_1|=\frac{\sqrt{(s-m_\pi^2-m_K^2)^2-4m_\pi^2m_K^2}}{2\sqrt{s}},
\label{eq:BG}
\end{eqnarray}
$|{\bf p}_1|$ being the CM momentum of the $\pi K$ system.

Each of $S^{(R)}_{11}$ is given by a corresponding amplitude 
$a_{11}^{(R)}$ taken as a simple relativistic Breit-Wigner form 
\begin{eqnarray}
S^{(R)}_{11} = 1+2ia^{(R)}_{11},\ \ 
a^{(R)}_{11}=
\frac{-\sqrt{s}\Gamma_R^1(s)}{s-M_R^2+i\sqrt{s}\Gamma_R^{\rm tot}(s)},
\label{eq:BW}
\end{eqnarray}
where $\Gamma_R^{\rm tot}(s)$($\Gamma_R^1(s)$) is a total width
(partial width of channel 1) of the resonance $R$,
given by
\begin{eqnarray}
\Gamma_R^{\rm tot}(s) = \sum_{i=1}^{3}\Gamma_R^i(s);\ \ 
\Gamma_R^i(s)=\frac{\rho_i}{\sqrt{s}}g_{Ri}^2
=g_{Ri}^2|{\bf p}_i|/8\pi s\ \  (i=1,2,3).
\label{eq:Gamma}
\end{eqnarray}
Here $g_{Ri}$'s are coupling constants to the channel $i$ of resonance R,
and the CM momentum $|{\bf p}_i|$ for $i=2,3$ 
are defined in a similar way as in Eq.(\ref{eq:BG}).
Thus parameters to be used for the fit are totally nine, {\em i.e.} 
resonance masses $M_R$'s (R=$\kappa$, $K_0^*$), 
their coupling constants $g_{Ri}$'s 
(i=1,2,3), and repulsive core radius $r_{c0}^\frac{1}{2}$. 

\section{Mass and width of $\kappa$ and core radius}
\label{sec:kap}
A high statistics data of the reaction 
$K^- p\rightarrow K^- \pi^+ n$ was obtained with 11 GeV/c beam using
LASS spectrometer at SLAC.\cite{rf:LASS} \
Spherical harmonic moments were used to perform an
energy independent Partial Wave Analysis of the
$K^-\pi^+$ system from threshold to 2.6 GeV,
with $t$-dependent parametrization of the production amplitudes.
The obtained $K^- \pi^+$-scattering amplitudes are the sum of
I=1/2 and 3/2 components.
The I=1/2 S-wave amplitude $a_0^\frac{1}{2}$
was determined
by subtracting the I=3/2 ($K^+\pi^+$/$K^-\pi^-$) component,
obtained independently by another experiment at SLAC.\cite{rf:estab} \
Here, the ``overall'' phase was fixed by 
imposing elasticity constraint to the amplitude in the $m_{K\pi}$
region below 1.29 GeV.
We use this amplitude between $K\pi$ threshold and 1.6 GeV for the analysis. 

\begin{figure}[p]
 \epsfysize=12.5 cm
 \centerline{\epsffile{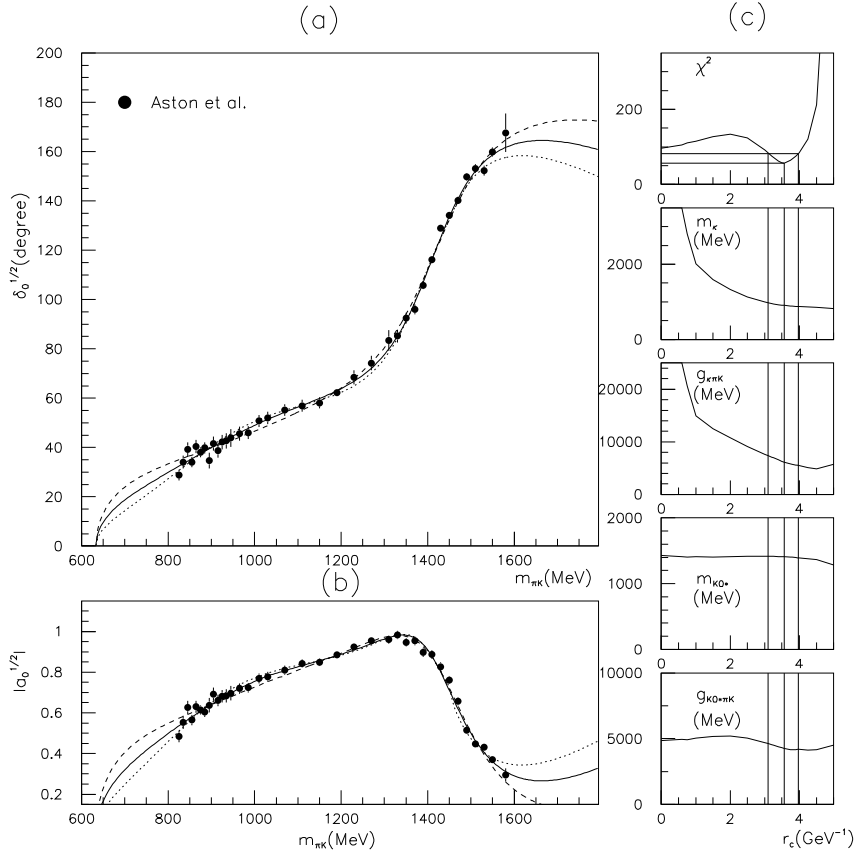}}
\caption{
Fits to I=1/2 $K\pi$ S-wave scattering amplitude; 
(a) phase shift $\delta_0^{1/2}$, and 
(b) magnitude of amplitude $|a_0^\frac{1}{2}|$.
The solid lines are the best fit with $r_c$=3.57GeV$^{-1}$, 
while the dotted and dashed lines are fits 
with $r_c$=3.1 and 3.975GeV$^{-1}$, respectively. 
(c) $\chi^2$, $M_\kappa$, $g_\kappa$, $M_{K_0^*}$, and $g_{K_0^*}$ behaviors 
as functions of core radius $r_c$. Vertical lines represent 
$r_c$=3.57, 3.1, and 3.975 GeV$^{-1}$, corresponding to the best fit 
and the fit with $\pm$ 5 sigma deviations.}
\label{fig:i12}
\end{figure}
\begin{table}[p]
\caption{Resonance parameters of $\kappa$(900), $K_0^*(1430)$ and core radius.
The errors correspond to five standard deviations from the best fit.
Two kinds of width, $\Gamma^{(p)}$ and $\Gamma^{(d)}$ defined as
$\Gamma^{(p)}$=$\Gamma_R^i(s=M^2)$(Eq.(5)),
$\Gamma^{(d)}$=$N^{-1}\int ds\Gamma (s)/[(s-M^2 )^2+s\Gamma (s)^2]$;
$N=ds\int 1/[(s-M^2 )^2+s\Gamma (s)^2]$, considering broadness of relevant
widths.
}
\begin{center}
\begin{tabular}{l|cccc}
\hline
\hline
 & $M_\kappa$     & $g_{K\pi}$  &
$\Gamma_{K\pi}^{(p)}$ &
$\Gamma_{K\pi}^{(d)}$ \\
\hline
$\kappa$(900) & 905$\stackrel{\scriptstyle +65}{\scriptstyle -30}$ MeV 
              & 6150$\stackrel{\scriptstyle +1200}{\scriptstyle -650}$ MeV
              & 545$\stackrel{\scriptstyle +235}{\scriptstyle -110}$ MeV 
              & 470$\stackrel{\scriptstyle +185}{\scriptstyle -90}$ MeV \\ 
$K_0^*$(1430) & 1410$\stackrel{\scriptstyle +10}{\scriptstyle -15}$ MeV 
              & 4250$\stackrel{\scriptstyle +380}{\scriptstyle -70}$ MeV
              & 220$\stackrel{\scriptstyle +40}{\scriptstyle - 5}$ MeV 
              & 220$\stackrel{\scriptstyle +40}{\scriptstyle - 5}$ MeV \\ 
\hline
\end{tabular}
\vspace*{5mm}
\begin{tabular}{c}
\hline
\hline
$r_{c0}^{1/2}$\\
\hline
3.57$\stackrel{\scriptstyle -0.45}{\scriptstyle +0.40}$GeV$^{-1}$ 
(0.70$\stackrel{\scriptstyle -0.09}{\scriptstyle +0.08}$fm)\\ 
\hline
\end{tabular}
\vspace*{5mm}
\label{tab:mw}
\end{center}
\end{table}
Figure~1(a) and (b)
show the result of the best fit to $\delta_0^\frac{1}{2}$
and $|a_0^\frac{1}{2}|$, respectively, by solid line.
The obtained parameters are collected in Table I.
The most remarkable feature is that we identify 
a low-mass resonance $\kappa$ with 
mass of about 900 MeV in the slowly-increasing phases 
between the threshold and 1300 MeV.
This is due to the role
of ``compensating" repulsive background $\delta_{BG}$,
whose existence is necessarily required from Chiral Symmetry
(see ii) and iii) of the following supplementary discussions).
As a matter of fact, the original LASS analysis of the data\cite{rf:LASS},
where a positive $\delta_{BG}$ with an effective range formulus 
was introduced, led to existence of only one state $K_0^*$(1430) 
with high mass. 
Since of the compensation between contributions due to $\kappa$ and  
the repulsive core, the mass value of $\kappa$ ($M_\kappa$) 
and its coupling to 
the $K\pi$ channel ($g_{\kappa 1}$)
are correlated to the core radius $r_c$.
To clarify this situation, various fits are performed 
with a series of fixed $r_c$ values
between 0 to 5.5 GeV$^{-1}$.
Fig.~1(c) shows the values of $\chi^2$, 
$M_\kappa$, $g_{\kappa 1}$ 
$M_{K_0^*}$, and  $g_{K_0^* 1}$ as functions of $r_c$.
$M_\kappa$ and $g_{\kappa 1}$ decrease as $r_c$ becomes larger,
while $M_{K_0^*}$ and  $g_{K_0^* 1}$ do not show such correlations
because these values are constrained mainly
by the steep phase increase around 1.4 GeV.
In the range of 2$\sim$5 GeV$^{-1}$, 
the $\chi^2$ value shows a parabolic shape, and 
makes its minimum at $r_c$=3.57 GeV$^{-1}$ where we get 
the best fit given in Fig.~1(a) and (b) ($\chi^2=57.0$ for 42 degrees  
of freedom; 51 data points with 9 parameters). 
When $r_c$ becomes smaller than 
2 GeV$^{-1}$, the values of $M_\kappa$ and $g_{\kappa 1}$ increase  
steeply, and the contribution of ``$\kappa$-meson resonance"  
has no more meaning than the positive background. 
A fit with $r_c$ setting to zero gives large  
$M_\kappa$ and $g_{\kappa 1}$ 
values (6.4 GeV and 39 GeV, respectively), and becomes essentially similar  
to the LASS analysis. This fit has the  
$\chi^2$ value of 96, which is larger  
by 40 than of our best fit.
\footnote{
In Ref.~\citen{rf:LASS} almost all data points are 
given with no accurate errors.
We will regard the original errors of $K^-\pi^+$ amplitude 
data\cite{rf:LASSd} equivalent to
our relevant errors of I=1/2 scattering amplitude,
which might be smaller than those of \citen{rf:LASS}.
This may be a reason why we get a $\chi$-square value larger than 
that in the LASS analysis.}

From the $\chi^2$ behavior in Fig.~1(c), 
we can obtain the upper and lower bounds of error 
of $M_\kappa$, $g_{\kappa 1}$ and $r_c$ whch are given in Table I as
five standard deviations from the best fit (+25 $\chi$-squares).
Corresponding curves with upper and lower values
of relevant parameters
are also shown in Fig.~1(a) and (b), respectively, 
by dotted ($r_c$=3.1 GeV$^{-1}$) and
dashed ($r_c$=3.975 GeV$^{-1}$) lines. 

The respective coupling constants to the 2nd channel 
$K\eta$ are obtained to be much smaller
than those to the $K\pi$ channel, {\em i.e.}, 
$g_{\kappa 2}$\raisebox{-0.5ex}{$\stackrel{<}{\sim}$} 1.0 GeV and 
$g_{K_0^* 2}$\raisebox{-0.5ex}{$\stackrel{<}{\sim}$} 0.9 GeV,
which are consistent to the elasticity constraint mentioned above.
The $g_{\kappa 3}$ and $g_{K_0^* 3}$, couplings to $K\eta'$ channel, 
are obtained with
much larger uncertainties, and their values are omitted here.

\section{Repulsive background in $K\pi$- and $\pi\pi$-systems}
\label{sec:rep}
In the present analysis, leading to the existence of $\kappa$-meson, 
introduction of a negative
background phase $\delta^{BG}$ of hard core type plays an essential role.
This is a similar situation as for $\sigma$-existence in $\pi\pi$-scattering.
In the I=2 $\pi\pi$ system, there are no known and/or expected resonances,
and this repulsive type phase shift itself is, if it exists, expected 
to be observed directly.
Actually a good fit to the experimental data was obtained\cite{rf:pl} 
by a similar formula of hard core type as Eq.(\ref{eq:BG})
with the core radius $r_{c0}^2=0.17$ fm.
The same situation is expected in the I=3/2 $K\pi$ scattering,
and we have made an similar analysis on the relevant 
$K\pi$-scattering phase shift.\cite{rf:estab}
The result is given in Fig.~2.
\begin{figure}[t]
 \epsfysize=5.0 cm
 \centerline{\epsffile{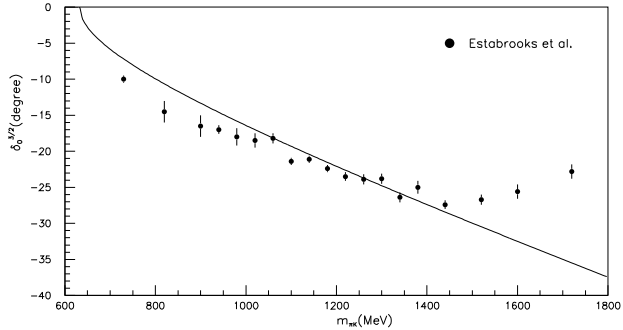}}
\caption{I=3/2 $K\pi$ scattering phase shift. Fitting by
hard core formula is also shown.}
\label{fig:i3/2}
\end{figure}
\begin{table}[t]
\caption{ Phenomenological core radii $r_{c0}^I$ in 
$\pi\pi$ and $K\pi$ systems}
\begin{center}
\begin{tabular}{l|cc|cc}
\hline
\hline
 &$(\pi\pi )_{I=0}$  
 &$(K\pi )_{I=1/2}$ &$(\pi\pi )_{I=2}$ &$(K\pi )_{I=3/2}$ \\
\hline
$r_c$ & 0.60$\pm$0.07fm & 0.70$\pm$0.09fm & 0.17fm &0.16fm \\ 
\hline
\end{tabular}
\vspace*{5mm}
\label{tab:hc}
\end{center}
\end{table}
The best fit is obtained with
the core radius $r_{c0}^{3/2}=0.16$ fm,
although the fit is somewhat worse than 
in the case of I=2 $\pi\pi$ system.

The values of phenomenological core radii in the $\pi\pi$ and $K\pi$ systems
are collected in Table II.
It is quite interesting that they are almost same within
the non-exotic channels (that is, $r^0$=$r^{1/2}$) and
within the exotic channels ($r^2$=$r^{3/2}$), respectively.
It seems to be reasonable from the viewpoint of $SU(3)$ flavor symmetry.

\newpage 
\section{Supplementary discussions}\label{sec:sup}
Here we give some additional comments on the results of our analysis:\\
i)
It may be interesting and important to compare
the properties of $\kappa$-meson obtained above
with predictions of the various theoretical models.
The SU(3) Linear $\sigma$ Model 
with the $U_A(1)$ breaking term(L$\sigma$M1)\cite{rf:gg,rf:shaba,rf:mss},
the SU(3) Linear $\sigma$ Model
without it(L$\sigma$M2)\cite{rf:sca}, 
and the extended Nambu Jona-Lasinio type 
model (ENJLM)\cite{rf:takizawa,rf:wise} (including also the $U_A(1)$
breaking term) give the $\kappa$-masses, respectively,
as $m_\kappa =1.2$\cite{rf:shaba},
0.8\cite{rf:sca} and
0.4$\sim$0.9\cite{rf:takizawa}
\footnote{This value is 
quoted in their analysis of case A, where $m_\sigma =0.604$GeV, 
close to our value\cite{rf:pipip,rf:pl}.
} in GeV, which have a large uncertainty.
The values of $\kappa$ decay width are given to be 1.2(0.1) GeV
in L$\sigma$M1 (L$\sigma$M2). The properties of $\kappa$ meson
in Table \ref{tab:mw}
seem not inconsistent with those predicted 
by the SU(3)-theoretical models with the 
$U_A(1)$ breaking term(L$\sigma$M1 and ENJLM). 
Accordingly our $\kappa$ meson
can be regarded as the member of $\sigma$-nonet, although
further investigations are necessary.\\
\ \ \      \\
ii) 
Concerning a possible origin of the repulsive background 
phase $\delta_{BG}$, introduced phenomenologically in this analysis, 
we should like to note its similarity\cite{rf:MY} to  
the $\lambda\phi^4$ term in L$\sigma$M. The $\lambda\phi^4$ term 
represents a strong repulsive and contact (zero-range) interaction 
between pions and seems to have a plausible property as an origin  
of $\delta_{BG}$, at least, in the low energy region, where the structures  
of composite pions may be neglected.

In $K\pi$($\pi\pi$)-scattering in L$\sigma$M
a contribution 
due to intermediate $\kappa$($\sigma$)-production in all $s$,$t$,$u$-channels
almost cancels\cite{rf:MY,rf:Tor} in the low energy region 
with a repulsive force from the $\lambda\phi^4$ interaction.
This leads effectively to the derivative (thus small) 
coupling\cite{rf:Ha} of Nambu-Goldstone
boson. This cancellation mechanism is guaranteed by chiral symmetry and PCAC.
It is notable that in the usual Breit-Wigner formula of S-wave resonance, 
a non-derivative
coupling of $\kappa$ and $\sigma$ resonance is supposed, without taking
the ``compensating" repulsive interaction into account.
This seems to be a reason why $\sigma$ and $\kappa$ resonances have been
overlooked in the many
phase shift analyses thus far made. 

The result given in Table II,
that the repulsive core radii in non-exotic channels are much 
larger than those 
in exotic channels, may be given\cite{rf:YITP}
some reason in L$\sigma$M as follows:
In exotic channels a large amount of strong repulsive force due to 
$\lambda\phi^4$ interaction 
is canceled by the attractive force due to crossed-channel exchange of relevant
scalar-mesons, while in non-exotic channels
there remains some amount of the repulsive force (going to compensate 
the attractive force due to $s$-channel intermediate production of 
the scalar mesons in the threshold).\\
\ \ \ \\
iii)
In analysis of the $\pi\pi$($K\pi$)-scattering, the importance
of $\rho$ ($K^*$) meson effects is often pointed out.\cite{rf:rho} 
In the S-wave scattering these vector mesons
contribute only through the crossed channel exchange diagrams, which
are necessarily accompanied by the ``compensating"
derivative $\phi^4$ interaction,\cite{rf:wein} \ similarly in the case of the
$\lambda\phi^4$ interaction to the $\kappa$ (or $\sigma$) exchange. 
They exactly cancel with each other
at the threshold and give only small effects 
in the low energy region, which may be regarded as included in the
background phase.\\
\ \ \ \\
iv)
Finally we give a comment on the behavior of the background phase.
In I=2 (I=3/2) channel of S-wave $\pi\pi(K\pi )$-scattering 
the fit of phase shifts by hard core formula
is satisfactory below $\sqrt{s}\sim 1.4$ GeV, as was shown in Fig.~1 
in Ref.~\citen{rf:pl}
(Fig.~\ref{fig:i3/2}).
However, the experimental phase shifts in the higher energy region seem to be 
decreasing\cite{rf:durusoy}. 
This is a very interesting phenomenon, which reminds us of the soft core
in nucleon-nucleon scattering. In this work we applied
the hard core type background, 
implicitly supposing ``local" $\pi$ and $K$ mesons.
The above mentioned ``soft core" type behavior of phase shifts 
in the comparatively high energy region
seems to suggest the composite structure of $\pi$ and $K$ 
as $q\bar{q}$-bound states. 

\begin{table}[t]
\caption{Candidate for Chiralons; members of chiral scalar-meson nonet.}
\begin{center}
\begin{tabular}{l|l|ccc|l}
\hline
\hline
 & & $M$(MeV)     & $g$(MeV)  & $\Gamma^{(p)}$(MeV) &  \\
\hline
I=0($n\overline{n}$) & $\sigma$(535-650) & 585$\pm$20 & 3600$\pm$350 
& 385$\pm$70  & $f_{1c}$ \\
I=0($s\overline{s}$) & $\sigma '$ & ?  & ? & ?  & $f^\prime_{1c}$ \\
I=1/2 & $\kappa$(900)     & 905$\pm$50 & 6150$\pm$900& 
 545$\pm$170 & $K_{1c}$ \\ 
I=1 & $\delta$  &    ?       &    ? & ?  & $a_{1c}$    \\ 
\hline
\end{tabular}
\vspace*{5mm}
\label{tab:ch}
\end{center}
\end{table}

\section{Concluding remarks}\label{sec:rem}
We have shown a strong evidence for existence of $\sigma$-meson 
in the previous work\cite{rf:pipip,rf:pl} and of $\kappa$-meson 
in the present work. The existence of these particles has a significant
importance in hadron spectroscopy. Since of their light masses
(and, for $\sigma$, of its vacuum quantum number),
they will appear in various processes such as 
$K\rightarrow 2\pi$ decay\cite{rf:MLS},
$K_{l4}$ decay\cite{rf:shaba2}, etc..

We have argued that $\kappa$
meson observed in the present analysis is a member of $\sigma$-meson nonet,
chiral partner of $\pi$-meson nonet. 
As was discussed in Ref.\citen{rf:pipip}, these particles 
(we call them ``Chiralons") should be
regarded as being different from the ordinary P-wave excited
states of $q\bar{q}$ system since of their light masses.
This discrimination may have also some theoretical reasons: 
In the extended
Nambu Jona-Lasinio model as a low energy effective theory of QCD
(, in which the existence of $\sigma$-meson is predicted),
only local composite quark and anti-quark operators
are treated, thus missing $L$-excited states in principle.
The present status of ``Chiralons,"\cite{rf:pipip}
is summarized in Table~\ref{tab:ch}.

Classification of low mass scalar mesons is still in confusion.
One of its main reasons 
seems to come from mis-identification of the chiral $\sigma$-nonet
with the $q\bar{q}$ $^3$P$_0$ nonet.
The properties of this extra-nonet should be further investigated 
also through the many other production
processes, such as $pp$-central collision\cite{rf:Taku},
$\Upsilon$ and $\Psi$ decays.  
In this connection especially the properties of observed resonances,
$f_0(980)$ and $a_0(980)$, are to be clarified in 
relation\cite{rf:shaba,rf:sca} to
the other members of chiralons, $\sigma^\prime$ with I=0 and $\delta$
with I=1. 

\newpage 
\section*{Acknowledgements}
The authors should like to express their deep gratitude to 
S.~Suzuki (Nagoya university) who has kindly informed us of 
Ref.~\citen{rf:LASSd}.

\end{document}